\documentstyle[11pt,draft]{article}
\newcommand{\be}{\begin{eqnarray}}
\newcommand{\ee}{\end{eqnarray}}
\newcommand{\bra}[1]{\mbox{$\langle\, #1 \mid$}}
\newcommand{\ket}[1]{\mbox{$\mid #1\,\rangle$}}

\newcommand{\expec}[1]{\mbox{$\langle\, #1\,\rangle$}}

\title{The accelerated observer with back-reaction effects}
\author{Roberto Casadio\thanks{e-mail: casadio@bo.infn.it} 
\ and Giovanni Venturi\thanks{e-mail: armitage@bo.infn.it}\\
 \\
{\em Dipartimento di Fisica, Universit\`a di
Bologna} \\
{\em and} \\
{\em Istituto Nazionale di Fisica Nucleare, 
Sezione di Bologna}\\
{\em via Irnerio 46, 40126 Bologna, Italy}}
\begin{document}
\begin{titlepage}
\pagestyle{empty}
\maketitle
\begin{abstract}
The quantum mechanical evolution of an accelerated extended detector coupled 
to a massless scalar field is exhibited and the back-reaction due 
to emission or absorption processes computed at first order in the change 
of the detector's mass and acceleration.
An analogy with black hole evaporation is found and illustrated.
\end{abstract}
\vskip 2truecm
\noindent
PACS numbers: 03.65.Sq, 04.70.Dy
\vskip 1truecm
\noindent
List of keywords:
\par
Accelerated observer\par
Back-reaction\par
Black hole evaporation\par
Two field detector\par
Unruh effect\par
\end{titlepage}
\pagestyle{plain}
\raggedbottom
\setcounter{equation}{0}
\section{Introduction}
\label{intro}
Because of the equivalence principle field theory in the presence
of gravitational fields is related to that in accelerated systems.
Indeed, subsequently to Hawking's remarkable discovery that black 
holes behave as if they had an effective temperature of 
$\hbar/8\,\pi\,G\,m$ with $m$ the mass of the black hole and $G$ 
Newton's constant \cite{hawking}, it was found that a detector with 
uniform acceleration in the usual vacuum state of flat Minkowski 
space will be thermally excited to a temperature 
$T=\hbar\,a/2\,\pi$ \cite{unruh}.
\par
It is therefore to be hoped that the study of the, apparently
simpler, Unruh effect may shed light on the case of a curved
space-time.
In particular, in his original paper \cite{unruh} Unruh suggested
a two-field model for a finite mass accelerated detector which
consisted of two scalars of differing mass.
This corresponds to a detector of finite mass having two energy levels 
separated by a gap corresponding to the mass difference of the two fields.
A point-like (infinite mass) monopole detector again having a
finite energy internal degree of freedom was suggested, on the other hand,
by DeWitt \cite{dewitt}.
It is clear that, if one considers a finite mass for the detector,
one must not ignore the quantum mechanical smearing of the trajectory
or the recoil back-reaction when a quantum is emitted (absorbed).
For the former reason we have previously \cite{cv} considered a massless
neutral scalar field $\varphi$ coupled to a finite mass quantum monopole
detector described by a gaussian wave-packet which was allowed to evolve
according to an inverted harmonic oscillator potential, corresponding
to constant acceleration.
On examining the probability per unit time that the detector be excited
by the absorption of scalar quanta, we observed that, on first 
considering the classical limit ($\hbar\to0$) and then the point-like 
limit for the gaussian wave-packet, we reproduced the usual 
results (Unruh effect \cite{unruh}).
If however one considered the point-like limit first, the detector 
decoupled from the scalar field.
This rather surprising result is associated with the fact that, once 
quantum-mechanical evolution is considered, the Compton wavelength of 
the detector enters in the theory.
\par
In our previous approach the scalar field only modified the internal
degree of freedom and did not influence the motion or mass of the 
detector.
The purpose of this letter is to include the back-reaction on the 
trajectory due to the emission (absorption) of massless scalar quanta
(our motivation is an analogy with Bremsstrahlung wherein a charged
particle decelerates by emitting soft photons and follows a
trajectory of changing energy and acceleration).
This will be done by modifying our previous approach in analogy 
with Unruh's two-field model in order to eliminate the need for the 
monopole moment (or internal degree of freedom) of the detector.
We shall then describe a ``detector'' with changing acceleration
and mass due to the emission (absorption) of scalar quanta which
will then mimic black hole evaporation thus generalizing the original
Unruh effect.
\par
We use units such that $c$ and the Boltzmann constant are equal 
to one.
\section{Two field model}
To illustrate our approach it will be sufficient (and easier) to consider 
a 2-dimensional Minkowski space-time with coordinates $x^0$, $z$ 
(for respectively Minkowski time and space) and an effective Lagrangian 
density
\be
{\cal L}(z,x^0)&=&\int d\tau\,\sum\limits_{i=1}^2\,\delta(x^0-x_i^0)
\nonumber \\
&&
\ \times
\left[i\,\hbar\,\psi^*_i(z,\tau)\,\dot\psi_i(z,\tau)
+{\hbar^2\over 4\,m_i}\,\left|{\partial\over\partial z}\psi_i(z,\tau)
\right|^2-m_i\,a_i^2\,z_i^2\,|\psi_i(z,\tau)|^2\right]
\nonumber \\
&&+{1\over 2}\,\int d\tau\,\delta\left(x^0-{1\over 2}\,(x^0_1+x^0_2)
\right)\,Q\,\left[\psi_2^*(z,\tau)\,\psi_1(z,\tau)
+\psi_2(z,\tau)\,\psi_1^*(z,\tau)\right]\,\varphi(z,\tau)
\nonumber \\
&&-{1\over2}\,\eta^{\mu\nu}\,\partial_\mu\varphi\,
\partial_\nu\varphi
\ ,
\label{L}
\ee
where $x^0_i =a_i^{-1}\,\sinh a_i\tau$, $a_i$ (positive) is the proper 
acceleration in the $z$ direction and a dot denotes differentiation with
respect to the continuous proper time $\tau$ which parametrizes the
(semi)classical trajectory followed by the observer $\psi$.
The diverse $\psi$ could be regarded as different states of the observer
(or detector).
A few words on the origin of our Lagrangian are in order:
in the classical limit the first term in ${\cal L}$ corresponds to the
Lagrangian for a inverted harmonic oscillator, that is
\be
L_{cl}=-m\,\left(\dot z^2+a^2\,z^2\right)
\ ,
\ee
and the sign is chosen so that the corresponding Hamiltonian is equal to
the (positive) particle (detector) mass (the opposite choice is made in
\cite{brout});
the second term describes the interaction between the detector and
the scalar field $\varphi$ (whose Lagrangian is given by the last term).
Further if one considers $Q=Q(\tau)$ as an operator acting on
the Hilbert space of the detector's internal energy states, $a_1=a_2$, 
$m_1=m_2$ one needs only one field $\psi$ ($=\psi_1=\psi_2$) and our 
previous results are reproduced \cite{cv}.
Instead we shall consider $Q$ a time independent c-number (coupling 
constant) and therefore the interaction with a quantum $\varphi$ is 
associated with the transition $\psi_1\to\psi_2$ corresponding to a 
change (which we shall always consider to be small) of acceleration 
and/or mass of the detector.
\par
From the first term in ${\cal L}$ one obtains
\be
\psi_i(z,\tau)&=&\left({\beta_i\over i\,b\,\sqrt{\pi}}\right)^{1/2}\,
\left({1\over2\,b^2}-i\,\beta_i\,\cosh a_i\tau\right)^{-1/2}
\nonumber \\
&& \times
\exp\left\{i\,\beta_i\,\left[z^2\,\cosh a_i\tau+
{\alpha_i^2\over4\,b^2\,\beta_i^2}\,
\left(a_i^{-2}\,\cosh a_i\tau-2\,z\,a_i^{-1}-4\,z^2\,\beta_i^2\,b^4\,
\cosh a_i\tau\right)\right]\right\}
\nonumber \\
&&  \times
\exp\left\{-{\alpha_i^2\over2}\,
\left(z-a_i^{-1}\,\cosh a_i\tau\right)^2\right\}
\ ,
\label{gausst}
\ee
where $\beta_i\equiv-{m_i\,a_i\over\hbar\,\sinh a_i\tau}$ and
$\alpha_i\equiv{2\,b\,\beta_i\over 
(1+4\,b^4\,\beta_i^2\,\cosh^2 a_i\tau)^{1/2}}$.
This is a solution to the equation of motion (Schr\"odinger equation)
with a gaussian wave-packet $\psi_i(z,0)$ of width $b$ as initial 
condition.
Since, as we have previously observed \cite{cv}, the Unruh effect
is obtained in the semiclassical limit ($\hbar\to 0$, $\beta_i\to-\infty$ 
and $b$ finite), it will be sufficient to use
\be
\psi_i(z,\tau)&\simeq&
{1\over\left(b\,\sqrt{\pi}\,\cosh a_i\tau\right)^{1/2}}\,
\exp\left\{-{i\over\hbar}\,m_i\,a_i\,z^2\,\tanh a_i\tau
-{\left(z-a_i^{-1}\,\cosh a_i\tau\right)^2\over 2\,b^2\,\cosh^2 a_i\tau}
\right\}
\ .
\label{gauss}
\ee
The energy of the above wave function, Eq.~(\ref{gauss}), may be evaluated 
obtaining
\be
\expec{H_i}=m_i\,\left[
{\hbar^2\over 4\,m^2_i}\,\bra{\psi_i}{\partial^2\over\partial z^2}\ket{\psi_i}
+a_i^2\,\bra{\psi_i}z^2\ket{\psi_i}\right]
=m_i
\ ,
\label{H}
\ee
which coincides with the Hamiltonian computed along the classical
trajectory $z_i=a_i^{-1}\,\cosh a_i\tau$, agrees with the rest mass
of the detector and occurs in the imaginary part of the exponent
of $\psi_i$ for $\tau\to0$ independently of the value of $a_i$.
\par
Let us now consider the probability for the detector to make a
transition from $a_1=a-\delta a/2$, $m_1=m-\delta m/2$
to $a_2=a+\delta a/2$, $m_2=m+\delta m/2$ associated with a 
quantum $\varphi$ of energy $|\delta m|$
($\delta m<0$ for emission and $\delta m>0$ for absorption and we shall see 
later that $\delta a$ is related to $\delta m$).
On using the interaction Lagrangian density (second term in 
Eq.~(\ref{L})) one obtains
\be
P_{21}(\delta a,\delta m)&=&
{Q^2\over 4\,\hbar^2}\,\int_{\tau_1}^{\tau_2} d\tau\,
\int_{\tau_1}^{\tau_2} d\tau'\,\int dz\,\int dz'\,
\psi_1^*(z',\tau')\,\psi_2(z',\tau')\,
\psi_1(z,\tau)\,\psi_2^*(z,\tau)
\nonumber \\
&&\phantom{-{Q^2\over 4\,\hbar}}\times
\bra{0}\,\varphi(z,x^0_c)\,\varphi(z',{x^0_c}')\,\ket{0}
\nonumber \\
&=&-{Q^2\over 4\,\hbar^2}\,\int_0^L d(\tau-\tau_1)\,
\int_0^L d(\tau'-\tau'_1)\,\int dz\,\int dz'\,
{\psi_1'}^*\,\psi_2'\,\psi_1\,{\psi_2'}^*
\nonumber \\
&&\phantom{-{Q^2\over 4}}\times
{\hbar\over 4\,\pi}\,\ln\left[\left(z-z'\right)^2
-\left(x_c^0-{x_c^0}'-i\,\epsilon\right)^2\right]
\ ,
\label{P}
\ee
where $x^0_c(\tau)\equiv \left(x^0_1(\tau)+x^0_2(\tau)\right)/2$,
${x^0_c}'\equiv x_c^0(\tau')$, $\psi_i'\equiv\psi_i(z',\tau')$.
In evaluating $P_{21}$ it is convenient to examine the exponent
of $\psi_2^*\,\psi_1$,
\be
\ln\,\psi_2^*\,\psi_1&=&-{1\over 2\,b^2}\,\left[
{\left(z-a_1^{-1}\,\sinh a_1\tau\right)^2\over\cosh^2 a_1\tau}
+{\left(z-a_2^{-1}\,\sinh a_2\tau\right)^2\over\cosh^2 a_2\tau}\right]
\nonumber \\
&&+{i\over\hbar}\,z^2\,\left(m_2\,a_2\,\tanh a_2\tau
-m_1\,a_1\,\tanh a_1\tau\right)
\nonumber \\
&&-{1\over 2}\,\ln\left(b^2\,\pi\,\cosh a_1\tau\,\cosh a_2\tau\right)
\ .
\label{expo}
\ee
On first considering the real part of the above it is immediate to see
that $|\psi_2^*\,\psi_1|^2$ is peaked at 
$z_c=\left(a_1^{-1}\,\cosh a_1\tau+a_2^{-1}\,\cosh a_2\tau\right)$
up to ${\cal O}[(\delta a)^2]$ corrections.
The imaginary part of Eq.~(\ref{expo}) may then be evaluated at $z=z_c$,
finally obtaining
\be
{i\over 2\,\hbar\,a}\,\left[
\left(\delta m+m\,{\delta a\over a}\right)\,\sinh 2a\tau
+2\,m\,\delta a\,\tau\right]
\ ,
\label{2.7}
\ee
again up to higher order corrections in $\delta a$, $\delta m$ and we 
have assumed $\delta a\,\tau$ is small (this will constrain the coupling
constant -- we shall return to this).
\par
The above imaginary part of the exponent is associated with the change 
of energy between the final (2) and initial state (1), which, from 
Eq.~(\ref{H}), is expected to be $\delta m=m_2-m_1$.
Hence requiring that Eq.~(\ref{2.7}) be equal to $-i\,\delta m\,\tau/\hbar$ 
leads to
\be
\delta m=-m\,{\delta a\over a}
\ ,
\label{2.8}
\ee
or
\be
m={f\over a}
\ ,
\label{2.9}
\ee
where $f$ is a positive constant.
Thus on demanding consistency ({\em i.e.}, conservation of energy/momentum)
we have obtained a relationship between
mass and acceleration corresponding to the action of a constant force.
\par
The above approximations and Eq.~(\ref{2.8}) may be substituted into 
Eq.~(\ref{P}) and on introducing $\tau=T+t/2$, $\tau'=T-t/2$, $T'=T-\tau_1$ 
one obtains
\be
P_{21}(\delta a)&=&
-{Q^2\over 8\,\pi\,\hbar}\,
\int_0^L dT'\,\int_{-L+2|T'-L/2|}^{+L-2|T'-L/2|} dt\,
e^{i\,{f\,\delta a\over \hbar\,a^2}\,t}\,
\ln\left[{2\over a}\,\sinh\left({a\,t\over 2}-i\,\epsilon\right)\right]
\nonumber \\
&=&
-i\,{Q^2\,a^2\over 8\,\pi\,f\,\delta a}\,\int_0^L dT'\,\int_{-L'}^{+L'} dt\,
\sum\limits_{n=0}^{\infty}\,
{e^{i\,{f\,\delta a\over \hbar\,a^2}\,t}\over
\left(t-{2\,\pi\,i\,n\over a}-i\,\epsilon\right)}
\ ,
\label{P1}
\ee
which is of the desired form (see, {\em e.g.}, \cite{birrell})
and we have omitted an end point contribution
in the integration by parts \cite{takagi}.
One may evaluate the $t$ integral in Eq.~(\ref{P1}) by closing the contour 
in the upper complex half plane ($\delta a>0$), thus including the poles 
at $t=2\,\pi\,i\,n/a$, with $n$ a non negative integer. 
One then obtains
\be
P_{21}(\delta a>0)\simeq
{Q^2\,a^2\,L\over 4\,f\,\delta a}\,\sum\limits_{n=0}^\infty\,
e^{-{2\,\pi\,f\,\delta a\over\hbar\,a^3}\,n}
\ ,
\label{P2}
\ee
up to contour contributions associated with transient effects, 
the requirements for the neglect of which we shall return to afterwards.
For the case $\delta a<0$ one performs the $t$ integration in Eq.~(\ref{P1}) 
in the lower complex half plane.
From the residues at $t=-2\,\pi\,n\,i/a$ one obtains the same result
as Eq.~(\ref{P2}) with $\delta a$ replaced by $|\delta a|$ and the sum now 
runs from $1$ to $\infty$ since the pole in $0+i\,\epsilon$, which is 
responsible of spontaneous emission/absorption, is excluded.
Thus one finally obtains
\be
P_{21}(\delta a)\simeq
{Q^2\,a^2\,L\over 4\,f\,|\delta a|}\,\sum\limits_n^\infty\,
e^{-{2\,\pi\,f\,|\delta a|\over\hbar\,a^3}\,n}
=
{Q^2\,a^2\,L\over 4\,f\,|\delta a|}\,{\sigma(\delta a)\over
1-e^{-{2\,\pi\,f\,\delta a\over\hbar\,a^3}}}
\ ,
\label{P3}
\ee
where $\sigma(x)=+1$ for $x>0$ and $\sigma(x)=-1$ for $x<0$.
In the above one notes the appearance of the familiar Planck distribution 
factor and the usual Unruh temperature $\beta^{-1}=\hbar\,a/2\,\pi$ 
(remembering that $\delta m=-f\,\delta a/a^2$ is the change in energy) 
with $a=(a_1+a_2)/2$.
Further we observe that for $f=1/4\,G$, $\beta^{-1}=\hbar/8\,\pi\,G\,m$
which is the Hawking temperature for a black hole of mass $m$.
\section{Multiple emission/absorption and black hole analogy}
\label{multi}
In the previous Section we have obtained an expression for the probability 
of emission (or absorption) of a scalar quantum by the accelerated 
observer with the corresponding back-reaction.
In this section we illustrate a possible analogy with black holes.
\par
To start with, let us note that for a generic accelerated observer one
needs an external static source to produce the constant force $f$.
Instead, if one identifies a black hole with its horizon (whose
acceleration is equal to the surface gravity $f/m=1/4\,G\,m$ and 
which is where Hawking emission takes place), the source of the
force coincides with the black hole itself and thus one does not
have an external source (as is the case in our model Eq.~(\ref{L})).
Hence, a change in the mass associated with the horizon is also
a change in the strength (mass) $m$ of the source, while $f=1/4\,G$
is constant (which is a statement of the equivalence principle).
\par
One may proceed to calculate the change of energy (mass) per unit time 
\cite{weldon} (see \cite{page} for an analogous calculation for a
black hole) for the accelerated observer emitting a ``thermal'' scalar
field $\varphi$ (see Eq.~(\ref{P3})) due to the coupling in our Lagrangian
(\ref{L}).
In particular the average loss of mass per unit time will then be given
by
\be
{\expec{\delta m}\over L}\propto
-{Q^2\over 4}\,\int{d\omega\over e^{\beta\,\hbar\,\omega}-1}
=-{Q^2\,f\over 8\,\pi\,m}\,\int{dx\over e^x-1}
\ ,
\label{dm}
\ee
where we have identified $\hbar\,\omega=f\,\delta a/a^2$, 
$x=\beta\,\hbar\,\omega$ and we shall not concern ourselves with eventual 
infrared divergencies which can be handled in the usual way \cite{weldon}.
Since we want the mass to decrease for increasing time (emission),
corresponding to the Unruh vacuum for a black hole \cite{birrell},
on identifying (for $L$ sufficiently small) 
$\dot m\simeq\expec{\delta m}/L$ one obtains
\be
m(\tau)=m_0\,\left(1-{\tau\over \tau_d}\right)^{1/2}
\ ,
\ee
which, on setting $f=1/4\,G$,corresponds to the evaporation of a
2-dimensional black hole (in the Schwarzschild time $\tau$) with initial
($\tau=0$) mass $m_0$ and decay time $\tau_d\propto 4\,\pi\,m_0^2/f\,Q^2=
16\,\pi\,G\,m_0^2/Q^2$. 
Correspondingly
\be
a(\tau)={f\over m_0}\,\left(1-{\tau\over \tau_d}\right)^{-1/2}
\ .
\label{atau}
\ee
On using Eq.~(\ref{dm}) it is straightforward to see that 
$\beta\,\expec{\delta m}$ is constant.
Clearly this result depends on the expression we obtained for $P_{21}$,
in particular the presence of the $1/\hbar\,\omega$ factor in 
Eq.~(\ref{P3}) which in turn depends on the form of the interaction in 
Eq.~(\ref{L}).
\par
The dependence of $a$ (or $m$) on time describes the semiclassical 
trajectory that the accelerated observer (black hole) follows.
One may then imagine replacing the continuously changing acceleration 
$a(\tau)$ by a series of $N$ straight lines each associated with equal time 
intervals $L$ and the emission of a scalar quantum every time the line slope 
changes. 
More precisely one considers a trajectory beginning at $\tau_0=0$
for an accelerated detector having initial mass $m_0$ and ending with a final 
mass $m_N$ in a time interval $\tau_N-\tau_0=NL$ ($\tau_N\le \tau_d$) 
after emitting $N$ quanta.
Clearly $L$ must be sufficiently small so that one may reasonably approximate
$a(\tau)$ in the above fashion and higher order terms be negligible.
\par
According to the above one obtains for the probability of
emission of $N$ quanta in the interval $NL$
\be
P_N=\prod_{r=1}^N\,P_{r,r-1}
\simeq
\left({Q^2\,L\over 4}\right)^N\,\prod_{r=1}^N\,
{(\delta m_r)^{-1}\over e^{\beta_r\,\hbar\,\omega_r}-1}
\ ,
\label{PN1}
\ee
where of course $\delta m_r=m_r-m_{r-1}\equiv m(rL)-m((r-1)L)$ and 
$P_{r,r-1}$ is given by Eq.~(\ref{P3}) with $a=(a_r+a_{r-1})/2$ and 
$\delta a=a_r-a_{r-1}$.
A remarkable point is that with the time dependence for the trajectory
given in Eq.~(\ref{atau}) (evaporation) one has
$\beta_r\,\hbar\,\omega_r\simeq \beta\,\expec{\delta m}=\mbox{\rm constant}$ 
and obtains
\be
P_N\propto\left({4\,\pi\,m_0/f\over
e^{\strut\displaystyle{\beta}\,\expec{\delta m}}-1}\right)^N\,
\prod_{r=1}^N\,\left(1-{rL\over\tau_d}\right)^{1/2}
\ ,
\label{PN}
\ee
corresponding to a sequence of $N$ emissions at the most probable
frequencies, that is the ones for which the exponents
in the denominators of Eq.~(\ref{PN1}) are minimum.
\par
Let us conclude by illustrating the constraints for the approximate validity
of our approach.
The final expression in Eq.~(\ref{PN}) depends on the accelerating force
$f$, 
the decay time $\tau_d$ (related to the coupling constant $Q$), the initial 
condition $m_0$ and the parameter $L$.
The time interval $L$ is constrained by the requirements that 
$\delta a\,L\ll 1$ (see after Eq.~(\ref{2.7})) and that the contour 
contributions in Eq.~(\ref{P1}) are negligible, that is the exponent
$f\,\delta a\,L/\hbar\,a^2\gg 1$.
Thus one needs
\be
{\hbar\over|\dot m|}\ll L^2\ll {m^2\over f\,|\dot m|}
\ ,
\ee 
which implies $m^2/f\gg\hbar$ or (for the black hole analogy $f=1/4\,G$)
$m\gg \sqrt{\hbar/G}\equiv m_p$, the Planck mass.

\begin{thebibliography}{99}
%
\bibitem{hawking} 
S. W. Hawking, {\it Nature} {\bf 248} (1974) 30;
{\it Comm. Math. Phys.} {\bf 43} (1975) 199.
%
\bibitem{unruh} 
W. G. Unruh, {\it Phys. Rev.} D {\bf 14} (1976) 870.
%
\bibitem{dewitt} 
B. S. DeWitt in {\it General Relativity: an Einstein
Centenary Survey}, S. W. Hawking and W. Israel eds. 
(Cambridge University Press, Cambridge, England, 1979).
%
\bibitem{cv} 
R. Casadio and G. Venturi, {\it Phys. Lett.} A {\bf 199} (1995) 33.
%
\bibitem{brout}
R. Brout and Ph. Spindel, Nucl. Phys. B {\bf 348} (1991) 405;
R. Brout, Z. Phys. B {\bf 68} (1987) 339.
%
\bibitem{birrell}
N. D. Birrell and P.C.W. Davies, {\it Quantum fields in curved space}
(Cambridge University Press, Cambridge, England, 1982).
%
\bibitem{takagi}
S. Takagi, {\it Prog. Theor. Phys.}, {\bf 74} (1985) 142.
%
\bibitem{weldon}
H. A. Weldon, {\it Phys. Rev.} D {\bf 49} (1994) 1579.
%
\bibitem{page}
D. N. Page, {\it Phys. Rev.} D {\bf 13} (1976) 198.
%
\end{thebibliography}
\end{document}